\documentclass[12pt]{article} 
\usepackage{secdot}
\usepackage[sectionbib]{natbib}
\usepackage{array,epsfig,fancyhdr,rotating}
\usepackage[]{hyperref}  
\renewcommand{\theequation}{\thesection\arabic{equation}}

\usepackage{changepage}
\usepackage{booktabs}
\usepackage{algpseudocode}
\usepackage{amsmath}
\usepackage{amssymb}
\usepackage{amsfonts}
\usepackage{multirow}
\usepackage{amsthm}
\usepackage{setspace}
\usepackage{tikz}
\usetikzlibrary{shapes,decorations,arrows,calc,arrows.meta,fit,positioning}
\tikzset{
    -latex,auto,node distance =2 cm and 2 cm, on grid, very thick,
    state/.style ={ellipse, draw, minimum width = 1 cm},
    point/.style = {circle, draw, inner sep=0.04cm,fill,node contents={}},
    bidirected/.style={Latex-Latex,dashed},
    el/.style = {inner sep=2pt, align=left, sloped}
}

\newtheorem{corollary}{Corollary}

\theoremstyle{definition}

\newtheorem{example}{Example}

\usepackage{color}
\usepackage{multirow}
\usepackage{xcolor}
\usepackage{bm}
\usepackage{enumitem}
\newtheorem{thm}{Theorem}
\numberwithin{equation}{section}

\newcommand{\indep}{\perp \!\!\! \perp}

\def\T{{ \mathrm{\scriptscriptstyle T} }}

\begin{document}


\renewcommand{\baselinestretch}{2}

\markright{ \hbox{\footnotesize\rm Statistica Sinica
}\hfill\\[-13pt]
\hbox{\footnotesize\rm
}\hfill }

\markboth{\hfill{\footnotesize\rm FIRSTNAME1 LASTNAME1 AND FIRSTNAME2 LASTNAME2} \hfill}
{\hfill {\footnotesize\rm Causal effects with non-Gaussianity} \hfill}

\renewcommand{\thefootnote}{}
$\ $\par


\fontsize{12}{14pt plus.8pt minus .6pt}\selectfont \vspace{0.8pc}
\centerline{\large\bf Identifiability and estimation of causal effects with}
\vspace{2pt}
\centerline{\large\bf  non-Gaussianity and auxiliary covariates}
\vspace{.4cm}
\centerline{Kang Shuai$^{a}$, Shanshan Luo$^{b}$, Yue Zhang$^a$, Feng Xie$^b$ and  Yangbo He$^{a \ast}$\footnote{$\ast$correspondence to: heyb@math.pku.edu.cn}}
\vspace{.4cm}
\centerline{$^a$Peking University, $^b$Beijing Technology and Business University}
 \vspace{.55cm} \fontsize{9}{11.5pt plus.8pt minus.6pt}\selectfont


\begin{quotation}
\noindent {\it Abstract: Assessing causal effects in the presence of unmeasured confounding is challenging. Although auxiliary variables, such as instrumental variables, are commonly used to identify causal effects, they are often unavailable in practice due to stringent and untestable conditions. To address this issue, previous researches have utilized linear structural equation models to show that the causal effect is identifiable when noise variables of the treatment and outcome are both non-Gaussian. 
In this paper, we investigate the problem of identifying the causal effect using the auxiliary covariate and non-Gaussianity from the treatment. Our key idea is to characterize the impact of unmeasured confounders using an
observed covariate, assuming they are all Gaussian. We demonstrate that the causal effect can be identified using a measured covariate, and then extend the identification results to the multi-treatment setting. We further develop a simple estimation procedure for estimating causal effects and derive a $\sqrt{n}$-consistent estimator. Finally, we evaluate the performance of our estimator through simulation studies and apply our method to investigate the effect of the trade on income.}

\vspace{9pt}
\noindent {\it Key words and phrases:}
Auxiliary variable, Causal effects, Identification, Multiple treatments, Non-Gaussianity.    \par
\end{quotation}\par

\def\thefigure{\arabic{figure}}
\def\thetable{\arabic{table}}

\renewcommand{\theequation}{\thesection.\arabic{equation}}

\fontsize{12}{14pt plus.8pt minus .6pt}\selectfont


\section{Introduction}
\hspace{0.6cm} Identifying and estimating the causal effect of a treatment on an outcome is crucial in practice, as it provides insight into the effectiveness of a given intervention. However, the existence of unmeasured confounding may be a core issue in observational studies. A common assumption of causal inference using observational data is exchangeability, which requires that one has measured a sufficiently rich set of covariates~\citep{rosenbaum1983assessing}. This is often challenged because investigators usually cannot accurately learn the confounding mechanism from the measured covariates in real scenarios.



Numerous strategies have been proposed for addressing identification issue of causal effects under unmeasured confounding. With a valid instrumental variable (IV) that satisfies relevance, independence, and exclusion restriction assumption, the causal effect can be identified for a binary treatment~\citep{angrist1996identification} or a continuous treatment~\citep{kasy2014instrumental} under certain conditions. Proxy variables can also be used to identify the causal effect, and certain types of proxy variables, known as negative control variables, are leveraged for mitigating confounding bias in observational studies~\citep{lipsitch2010negative,kuroki2014measurement,miao2018identifying,tchetgen2024introduction}. Among them, \citet{miao2018identifying} provides sufficient conditions for nonparametric identification of the causal effect using double negative control variables. For an overview of proxy variable based identification strategies, we refer readers to~\citet{tchetgen2024introduction}.

However, auxiliary variables like instrumental variables or proxy variables, may be unavailable in real scenarios because of stringent assumptions. To address this, many researchers focus on employing valid instrumental variables from a candidate instrument set~\citep{kang2016instrumental,guo2018confidence,windmeijer2019use} or imposing additional nonlinear model assumptions when only an invalid IV is available~\citep{liu2022mendelian,sun2022selective}. But the identification issue without a valid IV cannot be properly tackled even under the linear model setting. In comparison, with linear non-Gaussian model structure, identifying causal order or causal effects is still possible in the absence of auxiliary variables. The work of~\citet{shimizu2006linear} performs LiNGAM (short for Linear Non-Gaussian Acyclic Model) analysis and shows   {the causal structure}  is identifiable using independent component analysis~\citep{stone2002independent}.  { {Then the paper from~\citet{hoyer2008causal1} further slightly relax the non-Gaussian assumptions of LiNGAM.}} Following this, ~\citet{hoyer2008estimation} extends LiNGAM to linear non-Gaussian settings with hidden variables and shows that {  {the causal order and the causal effect}} are both identifiable.
Following their promising work, many researchers have accomplished work for causal discovery problems~\citep{shimizu2009estimation,shimizu2011directlingam,entner2011discovering,chen2013causality,salehkaleybar2020learning,wang2020high}. These articles assume non-Gaussianity, where all noise variables are non-Gaussian, except for~\citet{chen2013causality}, who considers the Gaussian confounder setting. Meanwhile, the use of non-Gaussianity to learn causal effects under hidden endogeneity is a common approach in econometrics. For example, \citet{cattaneo2012optimal} investigates the optimal inference problem with an IV when error variables are non-Gaussian. The research from~\citet{park2012handling} proposed that if the outcome error is Gaussian and the treatment is non-Gaussian, then the identification can be achieved by modeling the joint distribution of the endogenous regressor and the error term as a Gaussian copula. All of these researches reveal the potential of non-Gaussianity for identifying causal effects.

Regarding multiple treatments, identification issue becomes more difficult. The research of \citet{wang2019blessings} proposes a deconfounder method for identifying multiple treatment effects. They establish a factor model for treatments by assuming treatments are conditionally independent given confounders and adjust confounding bias using the estimated confounder. Following their work, various studies, comments, and discussions have emerged in the literature regarding multiple treatments~\citep{d2019comment,imai2019comment, kong2022identifiability, miao2023identifying}. They consider identifying effects of multiple causally independent treatments using a factor model and have highlighted the necessity of using auxiliary variables or non-Gaussianity for identification. For instance, \citet{kong2022identifiability} provides identification of causal effects for a binary outcome using non-Gaussianity from an auxiliary latent variable in the binary choice outcome model. However, identifying the causal effect for a continuous outcome, especially when treatments can be causally correlated, remains a unsolved challenge.

In this paper, we establish the identification results of the causal effect using a measured covariate and the non-Gaussianity from the treatment. This provides a new perspective for identifying the causal effect, which suggests that finding some Gaussian covariate instead of a valid IV in a candidate instrument set may also work. Additionally, we show that our proposed method can be extended to handle scenarios involving multiple treatments, among which the minimum number of required Gaussian covariates equals the smaller dimension between the dimension of treatments and the dimension of unmeasured confounders. We finally develop a simple estimation procedure for calculating causal effects using our strategy and derive a $\sqrt{n}$-consistent estimator. The simulation results show that our estimator can provide favorable performance. We finally apply our method to study the effect of the trade on the income, revealing the similar conclusion as the previous researches using a valid IV.

The remainder sections are organized as follows. Section~\ref{fram} provides the framework of our setting, including model and identification results in Section~\ref{mode}, and the extension to multi-treatment scenario in Section~\ref{exten}. In Section~\ref{asymp}, we present a simple estimation process and the $\sqrt{n}$-consistent estimator. We evaluate the
finite-sample performance of the proposed estimator in Section~\ref{simu}. Then, in Section~\ref{app}, we demonstrate the proposed strategy using the cross-sectional data for the year 2019 from the World Bank. Section~\ref{con} concludes the paper and provides several notable future research topics. The appendix presents the proof for Theorem~\ref{thm1}. Additional proofs and simulation results can be found in the supplementary material.



\section{Framework}\label{fram}
\subsection{Model and Identification}\label{mode}
\hspace{0.6cm}
Firstly, we consider the single treatment case. Denote $A\in \mathbb{R}^1$ as a treatment, $Y\in \mathbb{R}^1$ an outcome, $Z\in \mathbb{R}^1$ an observed covariate, $U\in \mathbb{R}^t$ the vector of unmeasured confounders. For simplicity, we consider a single measured covariate, { {but the same conclusion holds by conditioning on other measured covariates implicitly}}. Without loss of generality, we assume $U, Z$ both have zero mean and unit variance. We propose the following model
\begin{subequations}
\begin{align}
 \label{model1a}   A &= \gamma Z +  \lambda^\T U + \varepsilon_A, \\
 \label{model1b}   Y &= \alpha A + \beta Z + s^\T U + \varepsilon_Y,
\end{align}
\end{subequations}
where $(Z,U)$ are Gaussian with $\mathrm{cov}(Z,U)=\xi^{\T}$, $\varepsilon_A$ is non-Gaussian with non-zero variance, and $\varepsilon_A \indep (Z,U)$. The noise term $\varepsilon_Y$ satisfies $E(\varepsilon_Y\mid A) = 0$. Note that we do not require any prior knowledge about the distribution of $\varepsilon_Y$, so whether $\varepsilon_Y$ follows Gaussian distribution does not change the following conclusion, which differs from the requirement of~\citet{hoyer2008estimation} or~\citet{park2012handling}. { {Also, the Gaussianity of $Z$ can be directly tested and if we admit the Gaussianity of $U$ and the linear model assumption, then the non-Gaussianity of $\varepsilon_A$ is equivalent to the non-Gaussianity of treatment $A$}}.

Figure~\ref{fig1} provides a graphical illustration of our model. According to the potential outcome framework~\citep{rubin2005causal}, $Y_a$ is denoted as the outcome for an individual under a potential treatment $A=a$. We maintain the causal consistency assumption throughout the paper, implying no interference between units and there is only one version of the potential outcome~\citep{rubin1980randomization}. The average causal effect of $A$ on $Y$ for a unit increase of $A$ can be defined as follows
$$
\mathrm{ACE}_{A\to Y} = E ( Y_{a+1} - Y_a ) = \alpha.
$$
The main task is to identify the causal parameter $\alpha$ from observational data of $(Z,A,Y)$. We summarize the identification results in the following theorem. 
\begin{figure}
    \centering
\begin{tikzpicture}[>=stealth, scale=1.5]
    \node[draw, circle] (Z) at (1, 1) {$Z$};
    \node[draw, circle] (A) at (2.5, 1) {$A$};
    \node[draw, circle] (Y) at (5, 1) {$Y$};
    \node[draw, circle, densely dashed] (U) at (3.5, 2.5) {$U$}; 
    \draw[->] (Z) -- (A);
    \draw[->] (Z) edge[bend right]  (Y);
    \draw[->] (A) -- node[above] {$\alpha$} (Y);
    \draw[->] (U) -- (A);
    \draw[->] (U) -- (Y);
    \draw[-] (U) -- (Z);
\end{tikzpicture}
    \caption{Causal diagram with an observed covariate $Z$, a vector of unmeasured confounders $U$, a treatment $A$ and an outcome $Y$.}
    \label{fig1}
\end{figure}

\begin{thm}\label{thm1}
Under model \eqref{model1a}-\eqref{model1b} and assuming $\mathrm{cov}(Z,A)\neq 0$, the effect of $A$ on $Y$, namely $\alpha$, is identifiable.
\end{thm}
The above theorem establishes the identifiability of the causal effect $\alpha$ when the known source of non-Gaussianity arises from $\varepsilon_A$. As previously demonstrated, this conclusion holds regardless of the distribution of $\varepsilon_Y$. Unlike LiNGAM analysis assuming all noise variables to be non-Gaussian{{, focusing on the  identification of causal order}}, we only require non-Gaussianity from $\varepsilon_A$ for identifying a causal effect $\alpha$, which also differs from~\citet{park2012handling} assuming outcome error to be Gaussian.

The non-Gaussianity of $\varepsilon_A$ ensures that $E(Z\mid A)$ is a nonlinear function of $A$. Furthermore, the Gaussianity of $(Z,U)$ implies that $E(U\mid A)$ is proportional to $E(Z\mid A)$, which enables us to characterize the impact of $U$ with $Z$ when conditioning on $A$. Intuitively, the Gaussianity helps us separate the effect of $A$ on $Y$ from the effect of $(Z,U)$ on $Y$. Moreover, the non-Gaussianity of $\varepsilon_A$ makes it possible to identify $\alpha$ through the linear independence of $A$ and $E(Z\mid A)$. { {More generally, we can also include the interaction terms and some observed non-Gaussian covariates into consideration. This can be summarized in the following two corollaries.


\begin{corollary} \label{interaction}
    Replacing the model~\eqref{model1b} by $Y=\alpha A + \beta Z + s^\T U + r ZA + w^\T UA + \varepsilon_Y$ and also assuming (i) $\mathrm{cov}(Z,A)\neq 0$, (ii) $(Z,U)$ are Gaussian, (iii) $\varepsilon_A$ is non-Gaussian with non-zero variance, and (iv) $(Z,U,\varepsilon_A,\varepsilon_Y)$ have zero mean, then the effect of $A$ on $Y$, namely $\alpha$, is identifiable. 
\end{corollary}

\begin{corollary} \label{covariates}
    Replacing the models~\eqref{model1a}-\eqref{model1b} by the following models with independent observed non-Gaussian covariates $\Tilde{Z} \indep (Z,U)$
    \begin{subequations}
\begin{align}
 \label{model1_a}   A &= \gamma Z + \Tilde{\gamma}^\T \Tilde{Z} +  \lambda^\T U + \varepsilon_A, \\
 \label{model1_b}   Y &= \alpha A + \beta Z + \Tilde{\beta}^\T \Tilde{Z} + s^\T U + \varepsilon_Y,
\end{align}
\end{subequations}
and also requiring the same distributional assumptions as Corollary~\ref{interaction}, then the effect of $A$ on $Y$, namely $\alpha$, is identifiable.
\end{corollary}

Note that we center $(Z,U,\varepsilon_A,\varepsilon_Y)$ to be of zero mean in order to ensure $\mathrm{ACE}_{A\to Y} = E(Y_{a+1} - Y_a) = \alpha$, so $\alpha$ still represents the corresponding causal effect. Also, interaction terms and observed non-Gaussian covariates can be simultaneously taken into consideration as long as we combine the two corollaries. Throughout the following parts of the paper, we focus on  models~\eqref{model1a}-\eqref{model1b} for better illustration.}} Actually, our identification result provides a new perspective for identifying causal effects in the presence of unmeasured confounding. For example, with a candidate instrument set, \citet{kang2016instrumental} establishes the identification result of average causal effect with a candidate instrument set by assuming that at least 50\% of the instruments are valid. However, our identification results demonstrate that the causal effect $\alpha$ can be identified using an observed Gaussian covariate under the assumptions that $\varepsilon_A$ is non-Gaussian and $U$ is Gaussian. This is sufficient for identification, even when all candidate instruments are invalid. In such a situation, an alternative is to perform normality tests in the candidate instrument set instead of verifying stringent and untestable conditions of instrumental variable. Furthermore, the measured covariate $Z$ can serve as an invalid negative control variable, eliminating the need to find two valid negative control variables. The detailed proof of Theorem~\ref{thm1} can be found in the Appendix. The following example provides a specific illustration of Theorem~\ref{thm1}.

\begin{example} \label{ex1}
Suppose $Z,U,\varepsilon_Y$ are three independent random variables following standard normal distribution, $\varepsilon_A \sim \mathrm{U}(-1,1)$ and $\gamma = \lambda = 1$, we know that
$$
E(Y\mid A) =  \alpha A + \beta E(Z\mid A) + s E(U \mid A) = \alpha A + \Tilde{\beta} E(Z\mid A),
$$
where $\Tilde{\beta}$ is some constant because $E(U\mid A)$ is proportional to $E(Z\mid A)$ according to the proof of Theorem~\ref{thm1}. We show the contradiction if $E(Z\mid A)$ is linear in $A$:
\begin{align*}
    E(Z\mid A) = k A \Rightarrow E\{ (Z - k A) f(A) \} = 0,~\forall  f(\cdot),
\end{align*}
which implies $k = E\{ Z f(A) \} / E\{ A f(A) \}$. So we have
\begin{align*}
     f(A) = A  \Rightarrow k = \dfrac{E( Z A )} { E( A^2 ) } = \dfrac{3}{7}, ~~~~ f(A) = A^3  \Rightarrow k = \dfrac{E( Z A^3 )} { E( A^4 ) } = \dfrac{35}{81}, 
\end{align*}
where $k$ takes different values for different $f(\cdot)$. This is a contradiction, implying that $E(Z\mid A)$ cannot be linear in $A$. Hence, $A$ and $E(Z\mid A)$ must be linearly independent, which implies that causal parameter $\alpha$ will be identifiable.
\end{example}
The above example illustrates the implication of non-Gaussianity from $\varepsilon_A$. Such kind of non-Gaussianity allows us to distinguish $A$ from $E(Z\mid A)$. The similar idea can also be naturally extended to the nonlinear model setting, as illustrated in the following corollary.

\begin{corollary} \label{cor}
    Replace the outcome model in~\eqref{model1b} by $Y= f(A) + \beta Z + s^\T U + \epsilon_Y$, then $f(\cdot)$ is identifiable if $(\epsilon_A,Z,U)$ is Gaussian and either one of the following conditions holds
    \begin{itemize}
        \item[$\mathrm{(i)}$] $ \lim_{a\to \infty} f(a)/ \lvert a \rvert = 0$;
        \item[$\mathrm{(ii)}$]  $\text{ f(A) does not include the linear term of } A$. 
    \end{itemize}
\end{corollary}
When we replace model~\eqref{model1b} by $Y=f(A) + \beta Z + s^\T U + \epsilon_Y$, the main task becomes separating $f(A)$ from $E(Z\mid A)$. A natural choice for simplifying the discussion is by assuming that $\epsilon_A,Z$ and $U$ are Gaussian. In this case, $E(Z\mid A)$ is linear in $A$ and hence, we only need to separate $f(A)$ from $A$. The first condition in the above corollary is the so-called sublinear growth condition derived from the recent nonlinear causal discovery research~\citep{li2024nonlinear}. This implies that causal effects are still identifiable under Gaussian graphical modeling framework with nonlinear causation. Intuitively, the philosophy of separating $f(A)$ from $A$ is consistent with that in Theorem~\ref{thm1} by separating $A$ from $E(Z\mid A)$, as demonstrated in the following example.

\begin{example} \label{ex2}
Assume $Z,U,\varepsilon_Y,\varepsilon_A$ are four independent random variables following standard normal distribution in model \eqref{model1a}-\eqref{model1b}, we replace the linear outcome model by the following nonlinear additive model
$$
Y = f(A) + \beta Z + s U + \varepsilon_Y.
$$
Taking conditional expectation with respect to $A$ gives
$$
E(Y\mid A) = f(A) + \beta E(Z\mid A) + s E(U\mid A) = f(A) + \psi A,
$$
where $\psi$ is some constant. The main task is to identify $f(\cdot)$, which requires an additional assumption on $f(\cdot)$, e.g., the sublinear growth assumption from~\citep{li2024nonlinear}. This assumption means $\lim_{a\to \infty} f(a)/ \lvert a \rvert = 0$ and $\psi$ can be intuitively regarded as the slope in outcome model when we see the tail of $A$. Once $\psi$ is identified, $f(\cdot)$ can be immediately identified and estimated nonparametrically. If the Taylor expansion of $f(A)$ does not include the linear term of $A$, it is still possible to identify $f(\cdot)$.
\end{example}
This example implies that generalizing to a nonlinear outcome model can also be achieved. However, for the sake of simplicity, we restrict our attention to the linear setting. 
Next, we extend our identification results to the multi-treatment setting.

\subsection{Extensions}\label{exten}
\hspace{0.6cm}
We now extend our identification results from the single treatment case to the multiple treatments setting. Here, we omit the discussion with interaction terms and some observed non-Gaussian covariates because the similar conclusions, like those in Corollaries~\ref{interaction} and~\ref{covariates}, can be easily derived for multi-treatment cases. Let $A\in \mathbb{R}^p,Z\in \mathbb{R}^l$ and $U\in \mathbb{R}^t$, we consider the following linear model
\begin{subequations}
\begin{align}
 \label{model2a}   A &= \Gamma Z +  \Lambda U + \varepsilon_A,  \\
  \label{model2b}  Y &= \alpha^\T A + \beta^\T Z + s^\T U + \varepsilon_Y,
\end{align}
\end{subequations}
where $ \Gamma \in \mathbb{R}^{p \times l}, \Lambda \in \mathbb{R}^{p \times t} $, and $(Z,U)$ are jointly Gaussian with $\mathrm{var}(Z) = I_l, \mathrm{var}(U) = I_t$, and $\mathrm{cov}(Z,U) = \Sigma$. We summarize our identification results for multiple treatments in the following theorem.

\begin{thm}\label{thm2}
Under model \eqref{model2a}-\eqref{model2b}, the direct effect of each $A_j$ on $Y$, namely $\alpha_j$, is identifiable if the following conditions hold
\begin{adjustwidth}{-2.5pt}{}
\begin{itemize}
    \item[$\mathrm{(i)}$] $A$ and $ E(Z\mid A)$ are linearly independent;
    \item[$\mathrm{(ii)}$] $\mathrm{rank}(\Gamma + \Lambda \Sigma^{\T},  \Gamma \Sigma + \Lambda  ) = \mathrm{rank} (\Gamma + \Lambda \Sigma^{\T} )$,
\end{itemize}
\end{adjustwidth}
{ {where $\mathrm{rank}(\Sigma_1,\Sigma_2)$ represents the rank of the matrix  consisting of   all columns of $\Sigma_1$ and $\Sigma_2$.}}
\end{thm}
The proof of Theorem~\ref{thm2} can be found in the supplementary material. This theorem provides sufficient conditions for identifying causal effects of multiple treatments simultaneously. In practice, the conditions outlined by this theorem are relatively mild.
\begin{adjustwidth}{-3pt}{}
\begin{itemize}
    \item Condition (i) actually requires $E(Z\mid A)$ to be nonlinear in $A$ in the single treatment case with $Z\in \mathbb{R}^1$, which is equivalent to non-Gaussianity of $\varepsilon_A$. Moreover, Condition (i) can be tested using observational data of $(Z,A)$.
    \item Condition (ii) is sufficient to guarantee that $E(U\mid A)$ can be expressed as the linear combination of $E(Z\mid A)$, which is equivalent to requiring $\mathrm{cov}(Z,A)\neq 0$ in the single treatment case. More generally, condition (ii) implies that the column space of $\Gamma \Sigma + \Lambda$ is subsumed in the column space of $ \Gamma  + \Lambda \Sigma^{\T}$, which demonstrates that the equation $\mathrm{cov}(U,A) = \Phi \cdot \mathrm{cov}(Z,A)$ has a solution for $\Phi$. Condition (ii) holds if $\mathrm{cov}(A,Z) = \Gamma + \Lambda \Sigma^{\T}$ is of full row rank, which can be verified using observational data of $(Z,A)$.
\end{itemize}
\end{adjustwidth} 
Therefore, by finding a proper vector of Gaussian covariates $Z$ that satisfies both the linear independence condition (i) and the rank condition (ii), we can identify the multiple treatment effects $\alpha$.

How many observed Gaussian covariates do we need for identification of $\alpha$? We regard $Z$ as some randomized variable for convenience, implying that $Z$ and $U$ are independent with $\Sigma = 0$. In this case, condition (ii) reduces to $\mathrm {rank}(\Gamma, \Lambda) = \mathrm{rank}(\Gamma)$. Consider the special case $\mathrm{rank}(\Lambda) = \min(p,t)$, then the dimension of $Z$ should be at least $\min(p,t)$. This demonstrates that the minimum number of 
required Gaussian covariates equals the smaller dimension between the dimension of treatments and the dimension of unmeasured confounders. Especially in the single treatment case, one observed Gaussian covariate is sufficient for identifying $\alpha$. Under the scalar unmeasured confounder setting with $t=1$, one-dimensional $Z$ will also be enough for identification when $\Gamma$ is proportional to $\Lambda$.

Unlike the multiple treatments setting considered in~\citet{wang2019blessings}, we do not assume that treatments are conditionally independent given confounders. Instead, we utilize a linear model with distributional assumptions that permits causal relationships among treatments, even if the causal structure of A is unknown. In the context of multiple causally independent treatments, \citet{kong2022identifiability} provides identification results for binary outcome by assuming a linear Gaussian model for treatments under scalar unmeasured confounder setting. They require an additional auxiliary latent variable to be non-Gaussian in the binary choice outcome model. Our identification results serve as a complement of~\citet{kong2022identifiability} for identifying effects of multiple treatments with a continuous outcome, while not restricting the treatment model to have a factor structure and not assuming the dimension of unmeasured confounder to be one. The following example gives an illustration of Theorem~\ref{thm2} when effects of two causally correlated treatments are simultaneously of interest.


\begin{figure}
    \centering
\begin{tikzpicture}[>=stealth, scale=1.5] 
    \node[draw, circle] (Z) at (1.1, 1.8) {$Z$};
    \node[draw, circle, densely dashed] (U) at (3.7, 3.1) {$U$};
    \node[draw, circle] (A1) at (2.8, 2.2) {$A_1$};
    \node[draw, circle] (A2) at (2.8, 1.4) {$A_2$};
    \node[draw, circle] (Y) at (4.8, 1) {$Y$}; 
    \draw[-] (Z) -- (U);
    \draw[->] (Z) -- (A1);
    \draw[->] (Z) -- (A2);
    \draw[->] (Z) edge[bend right] (Y);
    \draw[->] (A1) -- (A2);
    \draw[->] (A1) -- node[above] {$\alpha_1$} (Y);
    \draw[->] (A2) -- node[above] {$\alpha_2$} (Y);
    \draw[->] (U) -- (A1);
    \draw[->] (U) -- (A2);
    \draw[->] (U) -- (Y);
\end{tikzpicture}
   \caption{Causal diagram with observed covariates $Z=(Z_1,Z_2)$, unmeasured confounders $U$, two treatments $(A_1,A_2)$ and an outcome $Y$.}
    \label{fig2}
\end{figure}

\begin{example} \label{ex3}
    Suppose we have two treatments $A_1,A_2$, and four independent random variables $U, \varepsilon_Y$, and $Z=(Z_1, Z_2)$, all following standard normal distributions. Let $\varepsilon_{A_1}\sim \mathrm{U}(-1,1)$ and $\varepsilon_{A_2}\sim \mathrm{U}(-2,2)$. The corresponding model is as follows (Figure~\ref{fig2} gives a graphical illustration)
    \begin{align*}
      A_1 =& Z_1 + U + \varepsilon_{A_1},  \\
      A_2 =& A_1 + Z_2 + U + \varepsilon_{A_2}, \\
        Y =& \alpha_1 A_1 + \alpha_2 A_2 + \beta_1 Z_1 + \beta_2 Z_2 + s U + \varepsilon_Y,
    \end{align*}
where the covariance matrix $\mathrm{cov}(A,Z)$ is of full row rank with $\mathrm{cov}(A_1,Z_1)=\mathrm{cov}(A_2,Z_1)=\mathrm{cov}(A_2,Z_2)=1$ and $\mathrm{cov}(A_1,Z_2)=0$, so condition (ii) of Theorem~\ref{thm2} is satisfied. To verify condition (i), we need to prove if there exists $ x_1,x_2,x_3,x_4 \
$ subject to
$$
x_1 A_1 + x_2 A_2 + x_3 E(Z_1\mid A_1,A_2) + x_4 E(Z_2 \mid A_1,A_2) =0,
$$
then $x_1=x_2=x_3=x_4=0$. This immediately implies
$$
E\{ (x_1 A_1 + x_2 A_2 + x_3 Z_1 + x_4 Z_2) \cdot g(A_1,A_2)   \} = 0, \quad \forall g(\cdot).
$$
We take $g(A_1,A_2) \in \{ A_1, A_2, A_1^3, A_2^3 \}$ and obtain the following linear equations
$$
\begin{pmatrix}
    7/3    & 10/3   &1    &0 \\
    10/3   &31/3    &1    &1 \\
    121/5  &131/5   &7    &0 \\
    516/5  &1197/5  &31   &31
\end{pmatrix} 
\begin{pmatrix}
    x_1 \\
    x_2 \\
    x_3 \\
    x_4
\end{pmatrix}
=0
,
$$
which implies that the unique solution is $x_1=x_2=x_3=x_4=0$ because the determinant of coefficient matrix is $-636$, and both conditions of Theorem~\ref{thm2} are satisfied. Thus, the effects of $A=(A_1,A_2)^\T$ on $Y$, namely $\alpha=(\alpha_1,\alpha_2)$, are identifiable.
\end{example}
The above example demonstrates the implications of our identification results for two causally correlated treatments. Unlike existing methods, our approach does not require the treatment model to have a factor structure where treatments are conditionally independent given confounders. Moreover, our results can be viewed as an extension of~\citet{kong2022identifiability} from binary outcome to continuous outcome using non-Gaussianity. The measured covariates $Z$ can be treated as invalid instruments or proxy variables, 
and their distributional assumptions can be verified empirically. While extending our results to a nonlinear outcome model for multiple treatments is feasible, we omit it here for simplicity.

\section{Estimation and Asymptotics} \label{asymp}
\hspace{0.6cm}
Before proceeding with the estimation procedure, it is essential to check all the required conditions using the observational data of $(Z,A,Y)$. For instance, we may use the Anderson-Darling test~\citep{anderson1952asymptotic} to assess the Gaussianity of $Z$ and the non-Gaussianity of $A$. Additionally, depending on the number of treatments, we may need to verify the conditions in either Theorem~\ref{thm1} or Theorem~\ref{thm2}. In the single treatment case, verifying $\mathrm{cov}(A,Z)\neq 0$ suffices. The only untestable assumption is the Gaussianity of the unmeasured confounders $U$.

After obtaining all the testable analysis results, we can construct a two-step estimator. In the first step, we estimate the conditional expectation $E(Z\mid A)$ using either a parametric working model or standard nonparametric regression methods. Although parametric methods such as maximum likelihood are suitable for estimation, they are prone to model misspecification. Alternatively,
nonparametric methods or machine learning techniques, such as random forest and multi-layer perceptron, have gained wide popularity because of 
their robustness to model misspecification~\citep{bickel1993efficient,chernozhukov2022automatic}.

After obtaining the estimator $\hat{E}(Z\mid A)$ for $E(Z\mid A)$, the second step involves directly regressing $Y$ on $A$ and $\hat{E}(Z\mid A)$ to construct a consistent estimator $\hat{\alpha}$ for $\alpha$. The estimation performance of $\hat{\alpha}$ heavily relies on the precision of the conditional expectation estimator. The final estimator of $\alpha$ is given by the formula
$$
\hat{\alpha} = \hat{X}_{11} \bigg(  \dfrac{1}{n} \sum_{i=1}^n A_i Y_i \bigg) + \hat{X}_{12} \bigg\{  \dfrac{1}{n} \sum_{i=1}^n \hat{E}(Z\mid A_i) Y_i  \bigg\},
$$
where
$$
\hat{X} = 
\begin{pmatrix}
    \hat{X}_{11} & \hat{X}_{12} \\
    \hat{X}_{12}^{\T} & \hat{X}_{22}
\end{pmatrix}
=
\left[
\dfrac{1}{n} \sum_{i=1}^n 
\begin{Bmatrix}
    A_i \\
    \hat{E}(Z\mid A_i)
\end{Bmatrix}
\begin{Bmatrix}
    A_i \\
    \hat{E}(Z\mid A_i)
\end{Bmatrix}^\T
\right]^{-1}.
$$
The estimation process is summarized in the following procedure called EUNC (short for Estimation Using Non-Gaussianity and Covariates).
\begin{flushleft}
\textbf{Input}: $n$ $\mathrm{i.i.d.}$ data $\{ Z_i,A_i,Y_i \}_{i=1}^n$ \\
\textbf{Output}: Estimator $\hat{\alpha}$ for effects of $A$ on $Y$
\begin{algorithmic}[1] 
\Procedure{EUNC}{$\mathrm{input}$}
\State Center the data of $(Z,A,Y)$ to be of zero mean.
\State Test $H_0: Z \text{ is Gaussian},A \text{ is non-Gaussian and }\mathrm{cov}(A,Z) \text{ is of full row rank}$.
\If{$H_0$ is true}
\State Use nonparametric methods to obtain estimator $\hat{E}(Z\mid A)$.
\State \textbf{if} $A,\hat{E}(Z\mid A)$ are linearly correlated
\State  \qquad \textbf{return} FAIL
\State \textbf{end if}
\State Regress $Y$ on $A,\hat{E}(Z\mid A)$
\Else \textbf{ return }FAIL
\EndIf
\State \textbf{return} the coefficient in front of $A$ as $\hat{\alpha}$
\EndProcedure
\end{algorithmic}
\end{flushleft}
{ {The EUNC procedure can also accommodate the situations with additional interaction terms like Corollary~\ref{interaction} or some observed non-Gaussian covariates like Corollary~\ref{covariates}. If interaction terms $A U^\T$ or $A Z^\T$ is present, we can include $Z E(Z\mid A)^\T$ in the EUNC procedure. If some observed non-Gaussian covariates $\Tilde{Z}$ is present, we should regress $Y$ on $A,\Tilde{Z},\hat{E}(Z\mid A,\Tilde{Z})$ rather than $A,\hat{E}(Z\mid A)$ at the final step of the EUNC procedure.}} The estimator obtained from the above procedure is $\sqrt{n}$ consistent, which is guaranteed when the convergence rate of $\hat{E}(Z\mid A)$ is at least $n^{-1/4}$. This convergence rate can be achieved by a correctly specified parametric model~\citep{hansen1982large} or by using existing nonparametric techniques~\citep{chernozhukov2022automatic}. We summarize this in the following theorem, and the proof is given in the supplementary material.
\begin{thm} \label{thm3}
    Suppose we obtain a uniformly consistent estimator $\hat{E}(Z\mid A)$ satisfying
    $$
   \sup_{A\in \mathcal{A}} \lvert \hat{E}(Z\mid A) - E(Z\mid A) \rvert = o_p(n^{-1/4}),
    $$
where $\mathcal{A}$ is the support of $A$ and $n$ is the sample size, then $\hat{\alpha}$ is root-$n$ consistent for $\alpha$ in the following sense
$$
\sqrt{n} (\hat{\alpha} - \alpha) \overset{d}{\to} N(0,\Sigma_{\alpha}),
$$
where
\begin{gather*}
\Sigma_{\alpha} = \mathrm{var} \bigg\{ X_{11} AY + X_{12} E( Z\mid A) Y + \dfrac{\partial \alpha(\mu)}{\partial \mu} \xi  \bigg\},\\
\xi = \mathrm{vec}\{ AA^\T, E(Z\mid A) A^\T, E(Z\mid A) E(Z\mid A)^\T \}, ~~\mu = E(\xi),\\
\alpha(\mu) = X_{11} E(AY) + X_{12} E\{ E(Z\mid A) Y \},
\end{gather*}
and
$$
X = 
\begin{pmatrix}
    X_{11} & X_{12} \\
    X_{12}^{\T} & X_{22}
\end{pmatrix}
= E\left[
\begin{Bmatrix}
    A \\
    E(Z\mid A)
\end{Bmatrix}
\begin{Bmatrix}
    A \\
    E(Z\mid A)
\end{Bmatrix}^\T
\right]^{-1}.
$$
\end{thm}
Here, the notation $\mathrm{vec}(\cdot)$ represents the vectorization of the corresponding matrix. The asymptotical variance can be estimated by the sample variance with all expectation terms replaced by their sample means. In our simulation and application parts, we will implement the bootstrap procedure in order not to introduce complex variance expressions. 

\section{Simulation Study}\label{simu}
\hspace{0.6cm}
In this section, we conduct several experiments to evaluate the performance of the proposed estimator. We compare our approach with two stage least square method (2SLS).

We generate seven simulated datasets based on the model \eqref{model1a}-\eqref{model1b}. The non-Gaussian noise \(\varepsilon_A\) is drawn from exponential distribution with rate 0.1 while \( Z, U \) and \( \varepsilon_Y \) are all marginally following a standard normal distribution. The value of $\alpha$ will be fixed at 1, the values of $\lambda, s$ will be fixed at 0.5, while the values of $\gamma, \beta, \xi$ will vary. If the corresponding edge in Figure~\ref{fig1} is present, we will set the corresponding parameter to a nonzero number, i.e., $\gamma = \beta =1,\xi=0.5$; otherwise, it will be set to 0.

The purpose of these settings is to compare the performance of our proposed estimator with  the commonly used 2SLS estimator. Cases 1 and 2 evaluate the performance of our estimator in the valid instrumental variable (IV) setting, with Case 1 representing a strong IV scenario with $(\gamma,\beta,\xi)=(1,0,0)$, and  Case 2 representing a weak IV scenario with $(\gamma,\beta,\xi)=(0.01,0,0)$. 
Cases 3-7 are designed to compare the two estimators in the presence of invalid IVs due to direct effects ($Z\to Y$) or unmeasured confounding.  
For more information on all simulated scenarios, please refer to Table \ref{tab:0318res}.

We consider sample sizes of $n\in \{100, 300, 500\}$ and perform 300 Monte Carlo simulations for each scenario. We employ the gradient boosting method to estimate $E(Z\mid A)$~\citep{friedman2002stochastic}. The simulation results are summarized in Table \ref{tab:0318res}.

\begin{adjustwidth}{-4pt}{}
\begin{itemize} 
    \item For Cases 1-2, covariate $Z$ can be viewed as a valid instrumental variable (IV). In Case 1, both methods exhibit relatively small bias, but our method has smaller standard errors and more significant coverage probability. However, in the weak IV setting (Case 2), the 2SLS estimator becomes more unstable with increased variances and conservative coverage probability, as reported in previous literature~\citep{Stock2002Survey}.
    \item Cases 3-5 allow for the exclusion restriction or independence assumption of the IV to be violated. In such cases, our EUNC procedure still yields a consistent estimator, while 2SLS results in a relatively large bias and poor coverage probability. Thus, we can conclude that employing a possibly invalid IV using the 2SLS method will provide worse estimation results.
    \item In Cases 6 and 7, covariate $Z$ can be regarded as a non-differential and an outcome-inducing confounding proxy~\citep{miao2018identifying}. Though $A$ and $Z$ are conditionally independent, the requirement $\mathrm{cov}(Z, A)\neq 0$ in Theorem \ref{thm1} still holds due to the presence of latent confounders $U$. The EUNC method exhibits remarkable robustness against deviations from invalid IV estimation and achieves superior performance across all evaluation metrics. 
\end{itemize}
\end{adjustwidth}
We can see that the EUNC procedure always provides favorable estimation results even when all edges in Figure~\ref{fig1} are present. More simulation results including two treatments case and sensitivity analysis for the Gaussianity of $U$ are provided in  the supplementary material.

\begin{table}[!htbp]
\centering
\caption{Comparison results of EUNC procedure and 2SLS. Here \(\checkmark\) implies the corresponding edge exists and \(\times\) means the absence of the edge. SD represents the standard deviation and 95\% CP is the coverage proportion of the 95\% asymptotic confidence intervals. Results are averaged over 300 repeated experiments. The bias and SD have been multiplied by 1000.\\}
\label{tab:0318res}
\resizebox{\textwidth}{!}{
\begin{tabular}{cccccccccccccc}
\toprule
 & \multirow{2}{*}{\(Z\rightarrow A\)} & \multirow{2}{*}{\(Z\rightarrow Y\)} & \multirow{2}{*}{\(U\rightarrow Z\)} & \multirow{2}{*}{\begin{tabular}[c]{@{}c@{}}Sample\\ size\end{tabular}} & & \multicolumn{2}{c}{Bias} && \multicolumn{2}{c}{SD} & & \multicolumn{2}{c}{95\% CP} \\
\cmidrule(l){7-8}  \cmidrule(l){10-11}  \cmidrule(l){13-14}
 &  &  &  & & & EUNC & 2SLS & & EUNC & 2SLS & & EUNC & 2SLS \\ [3pt] \hline 
\multirow{3}{*}{Case 1} & \multirow{3}{*}{\(\checkmark\)} & \multirow{3}{*}{\(\times\)} & \multirow{3}{*}{\(\times\)} & 100 & & 0.8 & 35.2 & & 15.2 & 4396.1 & & 96.7\% & 97.7\% \\ 
 &  &  &  & 300 & & 0.5 & 28.6 & & 7.7 & 2799.4 & & 95.7\% & 99.0\% \\
 &  &  &  & 500 & & 0.4 & 29.0 & & 5.8 & 1449.2 & & 96.0\% & 98.3\% \\ [3pt] \hline
\multirow{3}{*}{Case 2} & \multirow{3}{*}{\begin{tabular}[c]{@{}c@{}}\(\checkmark\)\\ (weak)\end{tabular}} & \multirow{3}{*}{\(\times\)} & \multirow{3}{*}{\(\times\)} & 100 & &  3.1 & 57.7 & & 15.4 & 6169.1 & & 97.3\% & 92.7\% \\
 &  &  &  & 300 & & 2.3 & 18.5 & & 8.2 & 18612.8  & & 98.3\% & 94.3\% \\
 &  &  &  & 500 & & 2.3 & 150.8 & & 6.4 & 8043.3  & & 95.7\% & 93.7\% \\ [3pt] \hline 
\multirow{3}{*}{Case 3} & \multirow{3}{*}{\(\checkmark\)} & \multirow{3}{*}{\(\checkmark\)} & \multirow{3}{*}{\(\times\)} & 100 & & 6.4 & 2394.6 & & 20.7 & 44166.6 & & 94.0\% & 95.0\% \\
 &  &  &  & 300 & & 3.1 & 1420.3 & & 9.9 & 29372.0 & & 97.3\% & 93.7\% \\
 &  &  &  & 500 & & 1.5 & 1442.4 & & 6.9 & 23295.0 & & 95.3\% & 94.7\% \\ [3pt]\hline 
\multirow{3}{*}{Case 4} & \multirow{3}{*}{\(\checkmark\)} & \multirow{3}{*}{\(\times\)} & \multirow{3}{*}{\(\checkmark\)} & 100 & & 0.7 & 175.6 & & 14.9 & 7749.6 & & 97.7\% & 95.3\% \\
 &  &  &  & 300 & & 0.7 & 282.4  & & 7.5 & 6789.6 & & 94.3\% & 93.7\% \\
 &  &  &  & 500 & & 1.2 & 77.3 & & 5.7 & 6341.5 & & 94.7\% & 87.3\% \\[3pt]\hline 
\multirow{3}{*}{Case 5} & \multirow{3}{*}{\(\checkmark\)} & \multirow{3}{*}{\(\checkmark\)} & \multirow{3}{*}{\(\checkmark\)} & 100 & & 9.5 & 835.8 & & 22.4 & 185903.3 & & 96.7\% & 91.7\% \\
 &  &  &  & 300 & & 3.6 & 1148.0 & & 10.0 & 14379.8 & & 96.7\% & 86.7\% \\
 &  &  &  & 500 & & 2.4 & 1210.0 & & 7.1 & 81481.0  & & 97.0\% & 81.7\% \\ [3pt]\hline 
\multirow{3}{*}{Case 6} & \multirow{3}{*}{\(\times\)} & \multirow{3}{*}{\(\times\)} & \multirow{3}{*}{\(\checkmark\)} & 100 & & 1.7 & 1978.0 & & 15.1 & 8744.9 & & 97.0\% & 87.7\% \\
 &  &  &  & 300 & & 0.2 & 476.9 & & 8.4 & 19060.6 & & 97.7\% & 89.7\% \\
 &  &  &  & 500 & & 0.7 & 1017.0 & & 6.5 & 22641.7 & & 96.3\% & 91.0\% \\ [3pt]\hline 
\multirow{3}{*}{Case 7} & \multirow{3}{*}{\(\times\)} & \multirow{3}{*}{\(\checkmark\)} & \multirow{3}{*}{\(\checkmark\)} & 100 & & 1.8 & 713.7 & & 25.4 & 56002.6 & & 97.0\% & 89.0\% \\
 &  &  &  & 300 & & 1.6 & 82528.3 & & 14.6 & 111138.4 & & 97.3\% & 89.7\% \\
 &  &  &  & 500 & & 0.9 & 1605.3 & & 10.6 & 88177.2 & & 96.3\% & 89.7\% \\ \hline
\end{tabular}}
\end{table}

\section{Application}\label{app}
\hspace{0.6cm}
In this section, we apply our method to a real-world dataset investigating the impact of the trade on income. We prioritize the search for Gaussian covariates, rather than focusing on identifying valid instrumental variables from a candidate set like previous literatures~\citep{Frankel1999Trade,kukla2009economic,lin2024instrumental, fan2024endogenous}. One of the main challenges in estimating the effect of trade on income is the potential endogeneity due to unmeasured common confounders for trade and income. Based on the study of \citet{lin2024instrumental}, we consider a modified structural model between income and trade share as follows:
\begin{equation}
    \Tilde{Y}_i = b + \alpha A_i + \beta \Tilde{Z}_i + \varepsilon_i, 
\end{equation}
where \( \Tilde{Y}_i = \log(Y_i), \Tilde{Z}_i = \log(Z_i)  \), and $Y_i$ is income per capita in country \(i\), \(A_i\) is the share of international trade to GDP, \(Z_i\) is an observed covariate, and $\varepsilon_i$ is an endogenous error term. 

\begin{figure}[htbp]
    \centering
    \includegraphics[width=0.8\textwidth]{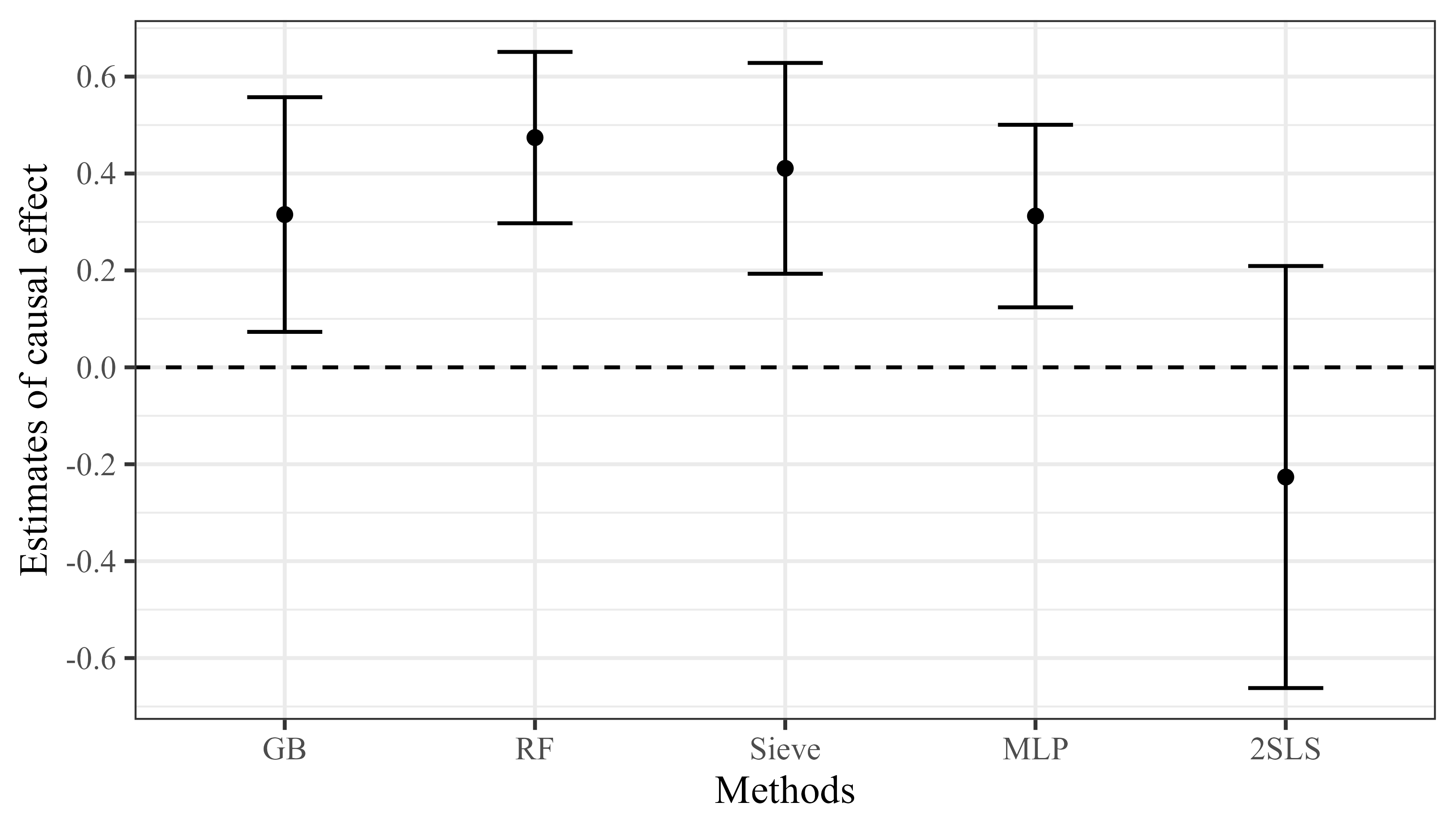}
    \caption{Confidence intervals of causal effect estimates for our method and 2SLS. Here GB, RF, Sieve and MLP represent Gradient Boosting, Random Forest, sieve estimation and Multi-Layer Perceptron. These methods are respectively used for calculating conditional expectation in our method.}
    \label{fig:annwat0.05}
\end{figure}

We collect cross-sectional data for the year 2019 from the World Bank and use 154 records for analysis after removing missing values. The trade share \(A\) is a non-Gaussian treatment because it fails to pass the Anderson-Darling test. We then perform the Anderson-Darling test to select an observed Gaussian covariate $\Tilde{Z}$, and the log value of total annual freshwater withdrawals (in billion cubic meters) is finally regarded as a Gaussian covariate with a \(p\)-value of 0.75.

The results are summarized in Figure~\ref{fig:annwat0.05}. We employ four nonparametric methods for calculating \( E( \Tilde{Z} \mid \Tilde{A} )\) and use 2SLS estimator as a comparison. The 95\% confidence intervals are obtained by nonparametric bootstrap method. Our method yields significantly positive effect estimates regardless of the  method used for estimating the conditional expectation. This corroborates with previous findings using instrumental variables~\citep{Frankel1999Trade,kukla2009economic,lin2024instrumental, fan2024endogenous}. In contrast, when considering the Gaussian covariate as a valid IV in the 2SLS method, there is no statistically significant evidence suggesting that the trade share affects income. These results indicate that our method can yield similar conclusions by employing a Gaussian covariate instead of a valid IV, which can be particularly useful when only potentially invalid instrumental variables are available. { {Practitioners can also include some interaction terms or observed non-Gaussian covariates into analysis  to enhance   credibility, as demonstrated by Corollary~\ref{interaction} and Corollary~\ref{covariates}.}}

\section{Conclusions and Future work}\label{con}
\hspace{0.6cm}
Our study focuses on the problem of identifying causal effects using auxiliary covariates and non-Gaussianity from the treatment. We present sufficient conditions for identifying effects of both single and multiple treatments using observed Gaussian covariates, which can be partially verified with observational data. Our results reveal that for a single treatment, identifying the causal effect can be facilitated by finding a Gaussian covariate { {with other non-Gaussian covariates being conditioned, rather than a valid instrumental variable.}} Additionally, our identification results for multiple treatments indicate that the minimum number of required Gaussian covariates equals the smaller dimension between the dimension of treatments and the dimension of latent confounders. Treatments can also be causally correlated without a factor structure.

There are several notable aspects of the approach developed in this paper. First, as illustrated in our second example, we can extend from a linear setting to a nonlinear outcome model. Second, we can perform sensitivity analysis to evaluate the robustness of Gaussian assumptions of latent confounders. Third, simulation studies show that our method has the potential to handle the weak instrumental variable problem. Thus, comparing theoretical properties between the proposed estimator and the 2SLS estimator would be an interesting problem. Fourth, efficiently searching for Gaussian covariates in a large candidate instrument set remains challenging but of great practical and theoretical interest. We leave the study of these issues as a future research topic as it is beyond the scope of this paper.

\section*{Supplementary Materials}
\hspace{0.6cm}
Supplementary Material available online includes  
 additional technical proofs and simulation results including two treatments case and sensitivity analysis of the Gaussianity of unmeasured confounders.

\section*{Acknowledgments}

 The research work is supported by National Key R\&D Program of China (2022ZD0160300). The authors thank the referees and an editor for helpful comments.

\section*{Appendix}\label{appendix}
\subsection*{Proof of Theorem~\ref{thm1}}
Let 
$$
\begin{pmatrix}
Z \\
U
\end{pmatrix}
\sim N \left\{ \mathbf{0},\begin{pmatrix}1 & \xi^\T \\ \xi & I_t \end{pmatrix} \right\},\quad \xi\in\mathbb{R}^{t}.
$$
The conditional expectation of $Z$ given $\varepsilon_A$ and $A$ will be
\begin{equation*}
    \begin{aligned}
    E(Z\mid \varepsilon_A, A)&=E(Z\mid \varepsilon_A, \gamma Z+\lambda^\T U+\varepsilon_A)\\
    &=E(Z\mid \varepsilon_A, \gamma Z+\lambda^\T U)\\
    &=E(Z\mid \gamma Z+\lambda^\T U),
    \end{aligned}
\end{equation*}
where the last equality holds because $\varepsilon_A \indep (Z,U)$. It is easy to observe
$$
\begin{pmatrix}
Z\\
\gamma Z+\lambda^\T U
\end{pmatrix}
\sim 
N\left\{ \mathbf{0},\begin{pmatrix}1 & \gamma+\xi^\T \lambda \\ \gamma +\lambda^\T \xi & \gamma^2+\lambda^\T \lambda + 2\gamma\xi^\T \lambda \end{pmatrix}\right\},
$$
which implies 
\begin{equation*}
\label{eq:cond_T}
    E(Z\mid \varepsilon_A, A)=E(Z\mid \gamma Z + \lambda^\T U) = \dfrac{\gamma + \xi^\T \lambda}{\gamma^2 + \lambda^\T \lambda + 2\gamma \xi^\T \lambda}(\gamma Z + \lambda^\T U).
\end{equation*}
Similar arguments show
\begin{equation*}
\label{eq:cond_U}
    E(U\mid \varepsilon_A, A) = E(U\mid \gamma Z + \lambda^\T U) = \dfrac{\gamma\xi + \lambda}{\gamma^2 + \lambda^\T \lambda + 2\gamma \xi^\T \lambda} (\gamma Z + \lambda^\T U).
\end{equation*}
When $\mathrm{cov}(A,Z) = \gamma + \xi^\T \lambda \neq 0$, we have
\begin{equation*}
\label{eq:rela_TU}
    E(U\mid \varepsilon_A, A)=\dfrac{\gamma\xi + \lambda}{\gamma + \xi^\T \lambda} E(Z \mid \varepsilon_A, A),
\end{equation*}
which means 
\begin{equation*}
\label{eq:rela_TU_1}
    E(U \mid A) = \dfrac{\gamma\xi + \lambda}{\gamma + \xi^\T \lambda} E(Z\mid A).
\end{equation*}
So the conditional expectation of $Y$ given $A$ will be 
\begin{equation}
\label{eq:identf_alpha}
    \begin{aligned}
     E(Y\mid A) & = \alpha A + \beta E(Z\mid A) + s^\T E(U\mid A)\\
                & = \alpha A + \bigg\{ \beta+\dfrac{s^\T (\gamma\xi + \lambda)}{\gamma + \xi^\T \lambda} \bigg\} E(Z\mid A).
    \end{aligned}
\end{equation}
Now we demonstrate that $E(Z\mid A)$ is nonlinear in $A$ when $\varepsilon_A$ is non-Gaussian with non-zero variance below. To show $E(Z\mid A)$ is actually nonlinear in $A$, we prove by contradiction. Suppose $E(Z\mid A) = c A$ for some constant $c\in \mathbb{R}^1$, then we must have
$$ E(U\mid A) = \dfrac{ \gamma\xi + \lambda}{\gamma + \xi^\T \lambda} E(Z\mid A) = \dfrac{c (\gamma\xi + \lambda)}{\gamma + \xi^\T \lambda} A, $$
which implies
\begin{align}
\label{cond}
E( \zeta\mid \zeta + \varepsilon_A ) = \gamma E(Z\mid A) + \lambda^\T E(U\mid A) = \Tilde{c} (\zeta + \varepsilon_A),
\end{align}
where
$$
A = \zeta + \varepsilon_A, \quad \zeta = \gamma Z + \lambda^\T U, \quad \Tilde{c} = c \bigg\{ \gamma + \lambda^\T \dfrac{ (\gamma\xi + \lambda)}{\gamma + \xi^\T \lambda} \bigg\}. 
$$
Then the equation~\eqref{cond} demonstrates 
\begin{align}
\label{cond_zero}
 E\{  ( 1- \Tilde{c}) \zeta - \Tilde{c} \varepsilon_A \mid \zeta + \varepsilon_A  \} =0.
\end{align}
Multiply equation~\eqref{cond_zero} by $\exp \{ it (\zeta + \varepsilon_A) \}$ and take expectation:
$$
E[  ( 1- \Tilde{c}) \zeta \exp \{it (\zeta + \varepsilon_A) \} ] -  E[ \Tilde{c} \varepsilon_A  \exp \{it (\zeta + \varepsilon_A) \} ] =0.
$$
The independence of $\zeta$ and $\varepsilon_A$ implies
$$
E\{  ( 1- \Tilde{c}) \zeta \exp ( it \zeta ) \} E\{ \exp (it \varepsilon_A ) \} -  E\{ \Tilde{c} \varepsilon_A  \exp ( it \varepsilon_A ) \} E \{ \exp ( it \zeta ) \} =0.
$$
Without loss of generality, we assume $\mathrm{var}(\zeta) = \mathrm{var}(\varepsilon_A) = 1$ below for convenience. The characteristic function of the standard normal distribution implies
$$
E \{ \exp ( it \zeta ) \} = \exp \bigg( -\dfrac{t^2}{2} \bigg), \quad E\{  ( 1- \Tilde{c}) \zeta \exp ( it \zeta ) \} = (1-\Tilde{c}) i t \exp \bigg(- \dfrac{t^2}{2} \bigg).
$$
Let $\phi(t) = E\{ \exp (it \varepsilon_A) \}$ and $\phi^{'}(t) = iE\{ \varepsilon_A \exp (it \varepsilon_A) \}$, then we have
$$
(1- \Tilde{c}) it \exp \bigg( - \dfrac{t^2}{2} \bigg) \cdot \phi(t) + \Tilde{c} \exp \bigg( - \dfrac{t^2}{2}  \bigg) \cdot i \phi^{'} (t) = 0,
$$
which means
\begin{align}
    (1- \Tilde{c}) t \phi(t) + \Tilde{c} \phi^{'}(t) = 0
\end{align}
When $\Tilde{c} \in \{ 0, 1\}$, $\phi(t)$ is a constant and $\varepsilon_A$ is deterministic, which is infeasible because we have assumed the variance of $\varepsilon_A$ is non-zero. Thus, for all $\Tilde{c} \not\in \{ 0, 1\}$, we have
\begin{align}
\label{diff_equation}
    \phi^{'}(t) + (\Tilde{c}^{-1} - 1)t\phi(t) =0, \text{ and } \phi(0) = 1.
\end{align}
Solving~\eqref{diff_equation} gives the characteristic function of $\varepsilon_A$
$$
\phi(t) = \exp \bigg(  - \dfrac{\Tilde{c}^{-1} - 1}{2} t^2 \bigg),
$$
which implies $\varepsilon_A$ must be Gaussian. This is a contradiction and hence, $E(Z\mid A)$ must be nonlinear in $A$. Now let 
$$
h =  \beta + \dfrac{s^\T (\gamma\xi + \lambda)}{\gamma + \xi^\T \lambda}, \quad g(A) = \big\{ A, E(Z\mid A) \big\}^\T.
$$ 
From equation~\eqref{eq:identf_alpha}, we have

$$
\begin{pmatrix}
\alpha \\
h
\end{pmatrix} =
E
\big\{
g(A) g(A)^\T
\big\}^{-1}
E\big\{ g(A) Y \big\},
$$
which means $\alpha$ is identifiable.



\begin{spacing}{1.25}

\bibhang=1.7pc
\bibsep=2pt
\fontsize{9}{14pt plus.8pt minus .6pt}\selectfont
\renewcommand\bibname{\large \bf References}
\expandafter\ifx\csname
natexlab\endcsname\relax\def\natexlab#1{#1}\fi
\expandafter\ifx\csname url\endcsname\relax
  \def\url#1{\texttt{#1}}\fi
\expandafter\ifx\csname urlprefix\endcsname\relax\def\urlprefix{URL}\fi

  \bibliographystyle{chicago}      
  \bibliography{ref.bib}   

\end{spacing}



%
%

\end{document}